\documentclass[twocolumn, 10pt]{bmc_article}

\usepackage{bm}%
\usepackage{times}
\usepackage{mathptm}
\usepackage{mysettings}
\usepackage{graphicx}
\usepackage{wrapfig}
\usepackage{amsmath,amssymb}
\usepackage{cite} % Make references as [1-4], not [1,2,3,4]
\usepackage{url}  % Formatting web addresses  
\usepackage[utf8]{inputenc} %unicode support
\allowdisplaybreaks
\begin{document}
%\begin{bmcformat}


\title{Effect of Peierls transition in armchair carbon nanotube \\ on dynamical behaviour of encapsulated fullerene}
 
%%%%%%%%%%%%%%%%%%%%%%%%%%%%%%%%%%%%%%%%%%%%%%
%%                                          %%
%% Enter the authors here                   %%
%%                                          %%
%% Ensure \and is entered between all but   %%
%% the last two authors. This will be       %%
%% replaced by a comma in the final article %%
%%                                          %%
%% Ensure there are no trailing spaces at   %% 
%% the ends of the lines                    %%        
%%                                          %%
%%%%%%%%%%%%%%%%%%%%%%%%%%%%%%%%%%%%%%%%%%%%%%

\author{Nikolai A Poklonski\correspondingauthor$^{1}$%
       \email{Nikolai A Poklonski\correspondingauthor - poklonski@bsu.by}%
      \and
       Sergey A Vyrko$^{1}$%
       \email{Sergey A Vyrko - vyrkosergey@tut.by}%
      \and
       Eugene F Kislyakov$^{1}$%
       \email{Eugene F Kislyakov - kislyakov@tut.by}%
      \and
       Nguyen Ngoc Hieu$^{2}$%
       \email{Nguyen Ngoc Hieu - hieunguyenvly@yahoo.com}%
      \and\\
       Oleg N Bubel'$^{1}$%
       \email{Oleg N Bubel' - poklonski@bsu.by}%
      \and
       Andrei M Popov\correspondingauthor$^3$%
       \email{Andrei M Popov\correspondingauthor - am-popov@isan.troitsk.ru}
      \and
       Yurii E Lozovik$^{3,6}$%
       \email{Yurii E Lozovik - lozovik@isan.troitsk.ru}
      \and
       Andrey A Knizhnik$^{4,5}$%
       \email{Andrey A Knizhnik - knizhnik@kintech.ru}
      \and\\
       Irina V Lebedeva$^{4,5,6}$%
       \email{Irina V Lebedeva - lebedeva@kintech.ru}
       and 
         Nguyen Ai Viet$^7$%
         \email{Nguyen Ai Viet - vieta@iop.vast.ac.vn}%
}

%%%%%%%%%%%%%%%%%%%%%%%%%%%%%%%%%%%%%%%%%%%%%%
%%                                          %%
%% Enter the authors' addresses here        %%
%%                                          %%
%%%%%%%%%%%%%%%%%%%%%%%%%%%%%%%%%%%%%%%%%%%%%%

\address{%
    \iid(1)Physics Department, Belarusian State University, pr. Nezavisimosti 4, Minsk 220030, Belarus\\
    \iid(2)North Carolina Central University, Durham, NC, 27707, USA\\
    \iid(3)Institute of Spectroscopy, Fizicheskaya Str. 5, Troitsk, Moscow Region, Russia, 142190\\
    \iid(4)RRC ``Kurchatov Institute'', Kurchatov Sq. 1, Moscow, Russia, 123182\\
    \iid(5)Kintech Lab Ltd, Kurchatov Sq. 1, Moscow, Russia, 123182\\
    \iid(6)Moscow Institute of Physics and Technology, Institutskii pereulok 9, Dolgoprudny, Moscow Region, Russia, 141701\\[-2.5pt]
    \iid(7)Institute of Physics and Electronics, Hanoi, Vietnam%
}%

\maketitle

%%%%%%%%%%%%%%%%%%%%%%%%%%%%%%%%%%%%%%%%%%%%%%
%%                                          %%
%% The Abstract begins here                 %%
%%                                          %%
%% The Section headings here are those for  %%
%% a Research article submitted to a        %%
%% BMC-Series journal.                      %%  
%%                                          %%
%% If your article is not of this type,     %%
%% then refer to the Instructions for       %%
%% authors on http://www.biomedcentral.com  %%
%% and change the section headings          %%
%% accordingly.                             %%   
%%                                          %%
%%%%%%%%%%%%%%%%%%%%%%%%%%%%%%%%%%%%%%%%%%%%%%
\vspace{-7mm}
\noindent
\begin{minipage}{\textwidth}
\begin{abstract}
The changes of dynamical behaviour of a single fullerene molecule inside an armchair carbon nanotube caused by the structural Peierls transition in the nanotube are considered. The structures of the smallest C$_{20}$ and Fe@C$_{20}$ fullerenes are computed using the spin-polarized density functional theory. Significant changes of the barriers for motion along the nanotube axis and rotation of these fullerenes inside the (8,8) nanotube are found at the Peierls transition. It is shown that the coefficients of translational and rotational diffusions of these fullerenes inside the nanotube change by several orders of magnitude. The possibility of inverse orientational melting, i.e. with a decrease of temperature, for the systems under consideration is predicted.
\end{abstract}
\end{minipage}

\vspace{-14mm}

%\ifthenelse{\boolean{publ}}{\begin{multicols}{2}}{}

%%%%%%%%%%%%%%%%%%%%%%%%%%%%%%%%%%%%%%%%%%%%%%
%%                                          %%
%% The Main Body begins here                %%
%%                                          %%
%% The Section headings here are those for  %%
%% a Research article submitted to a        %%
%% BMC-Series journal.                      %%  
%%                                          %%
%% If your article is not of this type,     %%
%% then refer to the instructions for       %%
%% authors on:                              %%
%% http://www.biomedcentral.com/info/authors%%
%% and change the section headings          %%
%% accordingly.                             %% 
%%                                          %%
%% See the Results and Discussion section   %%
%% for details on how to create sub-sections%%
%%                                          %%
%% use \cite{...} to cite references        %%
%%  \cite{koon} and                         %%
%%  \cite{oreg,khar,zvai,xjon,schn,pond}    %%
%%  \nocite{smith,marg,hunn,advi,koha,mouse}%%
%%                                          %%
%%%%%%%%%%%%%%%%%%%%%%%%%%%%%%%%%%%%%%%%%%%%%%

%%%%%%%%%%%%%%%%
%% Background %%
%%
\section{Introduction}
The structure and elastic properties of carbon nanotubes are
studied in connection with the perspectives of their
applications in nanoelectronic and nanoelectromechanical devices and
composite materials, and are also of fundamental interest,
particularly for physics of phase transitions. For example,
superconductivity \cite{Takesue06}, commensurate--incommensurate
phase transition in double-walled nanotubes~\cite{Bichoutskaia06},
spontaneous symmetry breaking with formation of corrugations along
nanotube axis~\cite{Connetable05} and structural Peierls transition
in armchair nanotubes~\cite{Mintmire92, Sedeki00, Viet94,
Harigaya93, Huang96, Poklonski08} have been considered. In the
present Letter, we consider a fundamentally new phenomenon related
to phase transitions in nanosystems. In other words, we
consider the possibility of inverse orientational melting for
molecules encapsulated inside nanotubes caused by structural Peierls
transition in the nanotubes.

\begin{figure}[!t]
\vspace{155mm}
\end{figure}

\begin{figure}%[!h]
\noindent\hfil\includegraphics[width=\columnwidth]{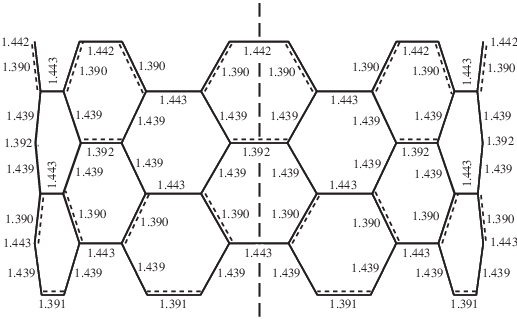}
\caption{{\bf Calculated Kekule structure corresponding to
the ground state of the (8,8) nanotube.} Nanotube axis is shown by the dashed line.}\label{fig:01}
\end{figure}

The possibility of Peierls transition in carbon nanotubes was first
considered in~\cite{Mintmire92}. As a result of this transition,
armchair nanotubes become semiconducting at low temperature, and
Peierls distortions lead to the Kekule structure (see
Figure~\ref{fig:01}) with two essentially different C--C bond
lengths and a triple translational period (three times more hexagons
in the translational unit cell). In previous studies, the Peierls
gap~\cite{Sedeki00, Viet94, Harigaya93} and the temperature of the
transition to the metallic phase with equal C--C bond
lengths~\cite{Mintmire92, Sedeki00, Huang96} were estimated.
Recently, the Kekule structure was calculated for the ground state
of an infinite armchair (5,5) nanotube by PM3 semiempirical
molecular orbital calculations~\cite{Poklonski08}. It was shown
that, for the (5,5) nanotube, the difference between C--C bond
lengths for semiconducting phase is 0.03~\AA, whereas the difference
between nonequivalent C--C bond lengths for metallic phase is only
0.006~\AA~\cite{Poklonski08}. Note that density functional theory
(DFT) calculations for the (5,5) nanotube of a finite length~\cite{Zhou04,
Matsuo03} also gave a 60 atom periodicity of physical properties on
the length of nanotube segment which is consistent with Kekule
structure for infinite armchair nanotubes. Moreover, X-ray
crystallographic analysis of chemically synthesized short
(5,5) nanotubes~\cite{Nakamura03} shows the Kekule bond length
alternation pattern, which was in good agreement with DFT and PM3 calculations also performed in~\cite{Nakamura03}. By the example of the
infinite (5,5) nanotube, it was demonstrated that the structural
Peierls transition connected with spontaneous symmetry breaking
takes place not only with an increase of temperature, but also can
be controlled by uniaxial deformation of armchair
nanotubes~\cite{Poklonski08}.

A dynamical behaviour of molecules encapsulated inside nanotubes can
correspond to the following regimes: oscillations about a fixed
position and/or a fixed orientation of the molecule (regime A),
hindered motion along the nanotube axis and/or rotation of the
molecule (regime B) and free motion and/or rotation of the molecule
(regime C). In the present Letter, we show the possibility of
changes of the dynamical behaviour of molecules encapsulated inside
armchair nanotubes as a result of the Peierls transition in the
nanotube structure. In other words, these changes can include
switching between the regimes A and B, switching between the regimes
B and C, and the changes in diffusion coefficients corresponding to
the hindered motion and/or rotation of the molecule (regime B). The
considered changes are possible in the case where the regime B takes
place for at least one phase of the nanotube, i.e. the temperature
$T_{\rm P}$ of the Peierls transition should correspond to the
temperature range of the regime B (the hindered motion and/or
rotation of the molecule). In other words, the temperature $T_{\rm
P}$ should be of the same order of magnitude (or a few orders of
magnitude less) as energy barriers $\Delta E$ for motion and/or
rotation of the molecule inside the nanotube at this phase. Note
that inverse melting of motion and/or rotation of the molecule is
possible, if the Peierls transition from the high- to
low-temperature phase of the nanotube occurs with switching from the
regime B to the regime C or switching from the regime A to the
regime B.

Estimations showed that the Peierls transition temperature is $T_{\rm
P} \simeq 1$--$15$~K~\cite{Mintmire92, Sedeki00, Huang96}. According
to calculations~\cite{Lozovik00, Lozovik02}, the barriers of the
value close to this temperature range were obtained for rotation of
the fullerene C$_{60}$ inside the C$_{60}$@C$_{240}$ nanoparticle.
It was also found that the changes of bond lengths of the fullerene C$_{240}$, the outer shell of these nanoparticles, within 0.06~\AA{}
lead to an increase of the barriers for rotation by more than an
order of magnitude~\cite{Lozovik00, Lozovik02}. The changes of the
nanotube bond lengths caused by the Peierls distortions are of the
same order of magnitude (about 0.02~\AA{} for the (5,5)
nanotube~\cite{Poklonski08}). Note also that the size of an
encapsulated molecule, and therefore, the nanotube radius cannot be
too large, since the magnitude of the Peierls distortions decreases
with an increase of the armchair nanotube radius~\cite{Viet94}.

Thus, taking into account the above considerations, we have chosen
the smallest fullerene C$_{20}$ and the magnetic endofullerene
Fe@C$_{20}$ to investigate changes in the dynamical behaviour of
molecules inside nanotubes at the Peierls transition. It has been
shown that the (8,8) nanotube is the smallest armchair carbon
nanotube which can encapsulate the fullerene C$_{20}$~\cite{Zhou06}.
A carbon nanotube with the fullerene C$_{20}$ inside was also used
as a model system to simulate a drug delivery via the
nanotube~\cite{Chen09}.

This Letter is organized as follows: Section 2 presents the DFT calculations of the structure of the C$_{20}$ and Fe@C$_{20}$ fullerenes and the PM3 calculations of the structure of the (8,8) nanotube. Section 3 presents the semiempirical calculations of the barriers for motion and rotation of the fullerenes inside the nanotube. Section 4 is devoted to the dynamical behaviour of molecules inside the nanotubes. Our conclusions are summarized in Section 5.

%%%%%%%%%%%%%%%%%%%%%%%%%%%%
%% Results and Discussion %%
%%
\begin{figure*}%[!h]
\noindent\hfil\includegraphics{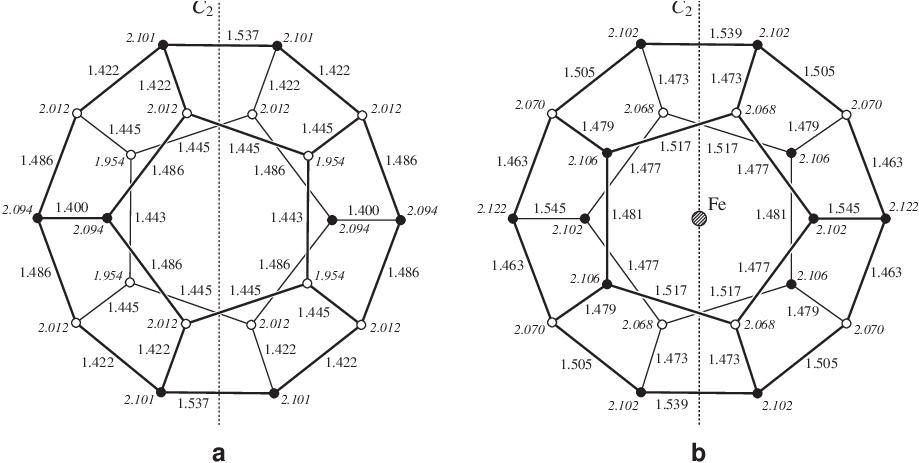}
\caption{{\bf Calculated structure of the C$_{20}$ (a) and
Fe@C$_{20}$ (b) fullerenes.}
The bond lengths are in \aa ngstr\"oms.
The distances between the carbon atoms and the fullerene centre (in
\aa ngstr\"oms) are denoted in italics. The atoms which have smaller
and greater distances to the fullerene centre are shown by the open
and filled circles, respectively. $C_2$ symmetry axis is shown by
the dotted line.}\label{fig:02}
\end{figure*}

\section{Fullerene and nanotube structures}
Structures of the C$_{20}$ and Fe@C$_{20}$ fullerenes have been
calculated using the spin-polarized density functional theory
implemented in NWChem 4.5 code~\cite{NWChem03} with the
Becke--Lee--Yang--Parr exchange--correlation functional
(B3LYP) \cite{Becke93, Lee88}. Eighteen inner electrons of the iron
atom are emulated with the help of the effective core
potential---CRENBS ECP \cite{Ross90} (only 8 valence
\emph{s}--\emph{d} electrons are taken into account explicitly).
The 6-31G* basis set is used for describing  electrons of the carbon
atoms.

The ground state of the fullerene C$_{20}$ is found to be a singlet
state and has $\bm{D}_{2h}$ symmetry. The calculated energy of the
triplet state of the fullerene C$_{20}$ is found to be 64 meV
greater than the energy of the ground state. The ground state of the
endofullerene Fe@C$_{20}$ is found to be a septet state and has
$\bm{C}_{2h}$ symmetry. The calculated structures of the ground
states of the C$_{20}$ and Fe@C$_{20}$ fullerenes are shown in
Figure~\ref{fig:02}. The iron atom locates in the centre of the
endofullerene. The smallest and the greatest distances between the
carbon atoms and the C$_{20}$ fullerene centre increase by 6 and
1\%, respectively, as a result of the iron atom encapsulation.

The semiempirical method of molecular orbitals modified for
one-dimensional periodic structures~\cite{Stewart87} with PM3
parameterization~\cite{Stewart89} of the Hamiltonian has been used
to calculate the structure of the (8,8) nanotube. The method was
used previously for calculating the Kekule structure of the (5,5)
nanotube ground state and for studying structural transitions
controlled by uniaxial deformation of this
nanotube~\cite{Poklonski08}. The adequacy of the PM3
parameterization of the Hamiltonian has been
demonstrated~\cite{Bubel00} by the calculation of bond lengths of
the C$_{60}$ fullerene with $\bm{I}_h$ symmetry: the calculated
values of the bond lengths agree with the measured
ones~\cite{Leclercq93} at the level of experimental accuracy of
$10^{-3}$~\AA. The calculated Kekule structure of the (8,8) nanotube
ground state is shown in Figure~\ref{fig:01}. The difference between
the lengths of short and long bonds of this Kekule structure of the
(8,8) nanotube is close to such a difference of the (5,5)
nanotube~\cite{Poklonski08}. The Peierls distortions include also
radial distortions of the armchair carbon nanotube with periodicity
of half of the translational period of the nanotube (for details
see~\cite{Poklonski08}). In the case of the (8,8) nanotube, the
longest nanotube radius is 0.547~nm, while the shortest radius is 0.544~nm.

\section{Fullerene--nanotube interaction}

The structures of the fullerenes C$_{20}$ and Fe@C$_{20}$ obtained in
section 2 have been used for finding
the ground state position and for studying motion and rotation of
these fullerenes inside the (8,8) nanotube. Two structures of the
(8,8) nanotube have been considered: the Kekule structure calculated
in section 2, and the structure
with all equal bonds 1.423 \AA~in length (so that to be equal to the
average bond length of the calculated the Kekule structure).

The calculations of the interaction energy between walls of
double-walled carbon nanotubes showed that the barriers for motion of
the short wall relative to the long wall are very sensitive to the
length of the long wall~\cite{Belikov04}. Thus, the size of the
system is too large for \emph{ab initio} calculations. The analogous
problem exists also for the considered case of a single fullerene
inside a nanotube. Therefore, the interaction between carbon atoms
of the fullerenes and the nanotube at the interatomic distance $r$
is described by the Lennard-Jones 12--6 potential

\begin{figure*}%[!h]
\noindent\hfil\includegraphics{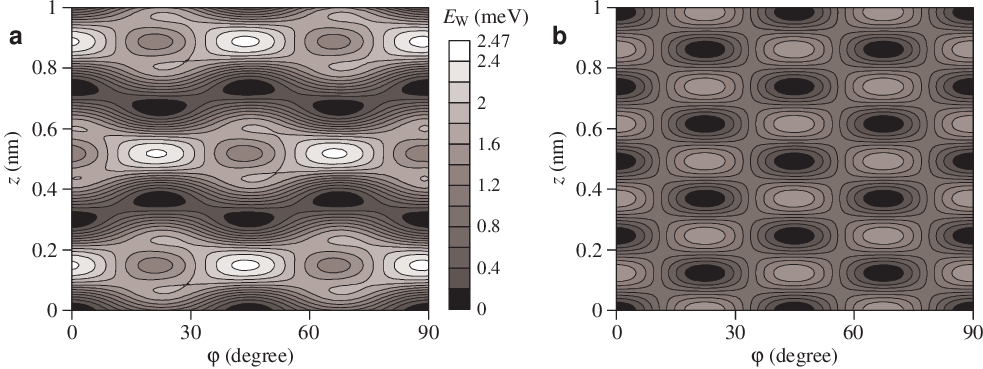}
\caption{{\bf The interaction energy $E_{\rm W}$ (in meV)
as a function of the fullerene C$_{20}$ displacement $z$ along the
nanotube axis and the angle $\varphi$ of the fullerene rotation
about the nanotube axis.}
{\bf (a)} The (8,8) carbon nanotube with
the Kekule structure; {\bf (b)} the (8,8) carbon nanotube with the
structure of metallic phase. The energy is given relative to the
energy minima. The equipotential lines are drawn at an interval
0.2~meV.}\label{fig:03}
\end{figure*}

\begin{equation}\label{eq:01}
   U_{\rm W} = 4\varepsilon \biggl(\biggl(\frac{\sigma}{r}\biggr)^{\!12} - \biggl(\frac{\sigma}{r}\biggr)^{\!6}\biggr)
\end{equation}
with the parameters $\varepsilon = 2.755$~meV, $\sigma = 3.452$~\AA.
These parameters of the Lennard-\hspace{0pt}Jones potential for the
fullerene--nanotube interaction are obtained as the average values
of the parameters~\cite{Girifalco00} for fullerene--fullerene and
fullerene--graphene interactions, in accordance with the procedure
described in~\cite{Girifalco00}. Here, the Lennard-\hspace{0pt}Jones
potential is used for calculating the potential surface of the
interaction energy $E_{\rm W}$ between the fullerene and the
infinite nanotube, and we believe that this gives adequate
qualitative characteristics of the potential surface shape. The
cut-off distance, $r = r_{\rm c}$ of the Lennard-Jones potential is
taken equal to $r_{\rm c} ={}$15~\AA. For this cut-off distance the
errors of calculation of the interaction energy $E_{\rm W}$ between
the fullerenes and the (8,8) nanotube and the barriers for relative
motion and rotation of the fullerenes inside the nanotube are less
than 0.1\%. Both the fullerenes and the nanotube are considered to
be rigid. An account of structure deformation is not essential for
the shape of the potential surface both for the interwall
interaction of carbon nanotubes~\cite{Kolmogorov00, Belikov04} and
the intershell interaction of carbon nanoparticles~\cite{Lozovik00,
Lozovik02}. For example, the account of the structure deformation of
the shells of C$_{60}$@C$_{240}$ nanoparticle gives rise to changes
of the barriers for relative rotation of the shells which are less
than 1\%~\cite{Lozovik00, Lozovik02}. It should also be noted that the
symmetry of interaction energy as a function of coordinates
describing relative positions of interacting objects is determined
unambiguously by symmetries of the isolated objects and does not
change if the symmetries of the objects are broken because of their
interactions.

The ground state interaction energies between the C$_{20}$ and
Fe@C$_{20}$ fullerenes, and the (8,8) nanotube with Kekule structure
are found to be $-$1.596 and $-$1.598~eV, respectively. The angles
between the $C_2$ symmetry axes of the C$_{20}$ and Fe@C$_{20}$
fullerenes and the nanotube axis at the ground states are
49.6$^{\circ}$ and 53.1$^{\circ}$, respectively. The metastable
states with the $C_2$ symmetry axes of the fullerenes perpendicular
to the nanotube axis are also found for both C$_{20}$ and
Fe@C$_{20}$. At the metastable states, the interaction energies are
greater by 4.58 and 3.04 meV than the ground state energies, for
C$_{20}$ and Fe@C$_{20}$, respectively.

\begin{figure*}%[!h]
\noindent\hfil\includegraphics{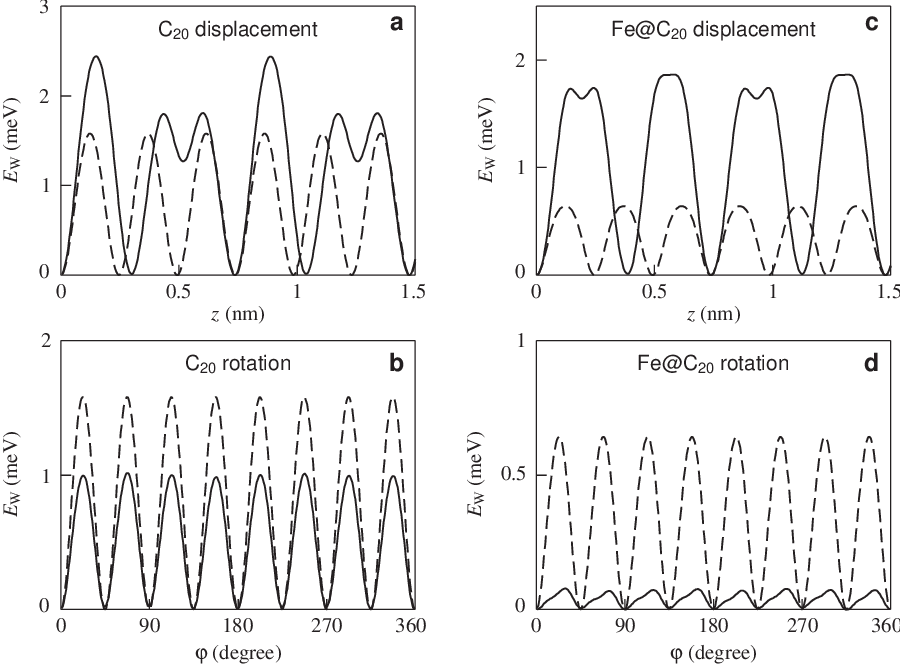}
\caption{{\bf Interaction energy between the C$_{20}$ fullerene or Fe@C$_{20}$ endofullerene and (8,8) nanotube.}
Dependences of the interaction energy $E_{\rm W}$ between the C$_{20}$ fullerene {\bf (a, b)} or Fe@C$_{20}$ endofullerene {\bf (c, d)} and the (8,8) nanotube on the fullerene displacement $z$ along the nanotube axis {\bf (a, c)} and on the angle $\varphi$ of the fullerene rotation about the nanotube axis {\bf (b, d)}. Solid lines denote nanotube with the Kekule structure, dashed lines denote nanotube with the structure of metallic phase. The energy minimum is positioned at $E_{\rm W}=0$, $z=0$ and $\varphi=0$.}\label{fig:04}

\bigskip\bigskip\medskip
\noindent\hfil\includegraphics{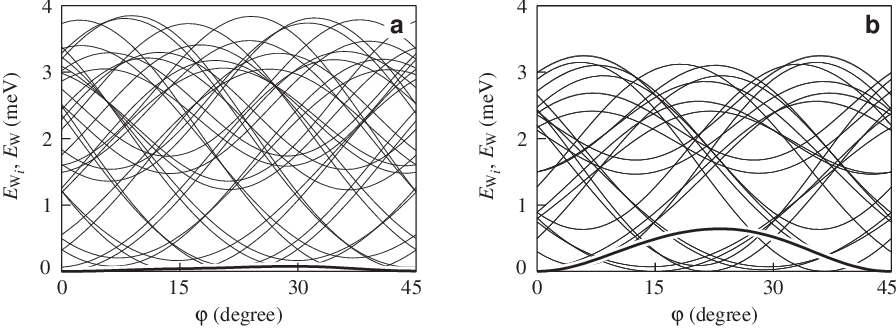}
\caption{{\bf Interaction energy between the Fe@C$_{20}$ endofullerene and (8,8) nanotube.}
Dependences of the interaction energy $E_{{\rm W}i}$ between the Fe@C$_{20}$ endofullerene and individual atoms of the (8,8) nanotube on the angle $\varphi$ of rotation of the endofullerene about the nanotube axis are denoted by the thin lines. Dependence of the total interaction energy $E_{\rm W}$ between the Fe@C$_{20}$ endofullerene and the (8,8) nanotube on the angle $\varphi$ is denoted by the thick line. {\bf (a)} The (8,8) carbon nanotube with the Kekule structure; {\bf (b)} the (8,8) carbon nanotube with the structure of metallic phase. All energies are given relative to the energy minima. Only dependences $E_{{\rm W}i}$ with high values of the barriers $\Delta E_{{\rm r}i}$ are shown.}\label{fig:05}
\end{figure*}

The potential surfaces of the interaction energy between the
C$_{20}$ fullerene and the nanotube, $E_{\rm W}(\varphi,z)$, as
functions of the relative displacement of the fullerene along the
axis of the nanotube $z$ and the angle of relative rotation of the
fullerene about the axis of the nanotube $\varphi$ are presented in
Figure~\ref{fig:03} for both the considered structures of the (8,8)
nanotube. In the general case, diffusion of a fullerene along the
nanotube axis is accompanied by rotation of the fullerene. Our
calculations show that the barriers for rotation of both fullerenes
about the axes which are perpendicular to the nanotube axis lie
between 3 and 23~meV for any orientation of the fullerene and for
both the considered structures of the (8,8) nanotube. These barriers
are significantly greater than the barriers for rotation of the
fullerenes about the axis of the nanotube (shown in
Figure~\ref{fig:03} for the C$_{20}$ fullerene). Therefore,
diffusion of the fullerenes along the nanotube axis is accompanied
only by rotation about the nanotube axis. Thus, the minimal barrier
$\Delta E_{\rm d}$ for diffusion of the fullerenes along the
nanotube axis is the barrier between adjacent minima of the
potential surface $E_{\rm W}(\varphi,z)$. Figure~\ref{fig:03} shows
that the shapes of the potential surface $E_{\rm W}(\varphi,z)$,
corresponding to the Kekule structure of the (8,8) nanotube and to
the structure of metallic phase are essentially different. For a
case of the structure corresponding to the metallic phase of the
(8,8) nanotube, all the barriers between adjacent minima of the
potential surface $E_{\rm W}(\varphi,z)$ are equivalent. In this
case, the same barrier $\Delta E_{\rm d} = \Delta E_{\rm r}$ should
be overcome for diffusion of the fullerenes along the nanotube axis
and for rotation of the fullerenes about this axis (see
Figure~\ref{fig:03}b). For a case of the Kekule structure, two
different barriers between the adjacent minima exist: the barrier
$\Delta E_{\rm d}$ for diffusion of the fullerenes along the
nanotube axis, and the barrier $\Delta E_{\rm r}$ to rotation of the
fullerenes about this axis (see Figure~\ref{fig:03}a). For the
Fe@C$_{20}$ fullerene, the shapes of the potential surfaces $E_{\rm
W}(\varphi,z)$ are qualitatively the same as for C$_{20}$ for both
the considered structures of the nanotube. The calculated values of
the barriers $\Delta E_{\rm d}$ and $\Delta E_{\rm r}$ are listed in
Table \ref{table1} for both fullerenes. The dependences of the
interaction energy between the C$_{20}$ and Fe@C$_{20}$ fullerenes
and the nanotube on the relative displacement of the fullerene along
the axis of the nanotube and the angle of its relative rotation
about this axis, corresponding to both the considered structures of
the (8,8) nanotube are compared as shown in Figure~\ref{fig:04}.
Figure~\ref{fig:04} is a vivid illustration of the significant
changes of the barriers $\Delta E_{\rm d}$ and $\Delta E_{\rm r}$.
The most dramatic change corresponds to the rotation of the
Fe@C$_{20}$ fullerene about the nanotube axis.

\begin{table}[!t]
\vspace{-6pt}
\caption{\bf Calculated characteristics of the dynamical behaviour of the C$_{20}$ and Fe@C$_{20}$ fullerenes inside the (8,8) nanotube of different structure}\label{table1}
{\small
\medskip
\begin{tabular}{lcccc} \hline
\bf \footnotesize Nanotube structure\hspace{-4.5pt} & \multicolumn{2}{c}{\bf \footnotesize Kekule structure} & \multicolumn{2}{c}{\bf \footnotesize Metallic phase structure} \\\hline 
\bf \footnotesize Fullerene &\bf \footnotesize C$_{20}$ &\bf \footnotesize Fe@C$_{20}$ &\bf \footnotesize C$_{20}$ &\bf \footnotesize Fe@C$_{20}$
\\\hline
\vline width0pt height10pt depth0pt $\Delta E_{\rm d}$ (meV) & 1.68 & 1.73 & 0.87 & 0.44 \\
$\Delta E_{\rm r}$ (meV) & 0.33 & 0.05 & 0.87 & 0.44 \\
$\nu_{\rm d}$ (GHz) & 72.2 & 54.8 & 75.7 & 51.3 \\
$\nu_z$ (THz) & 0.439 & 0.144 & 0.558 & 0.329 \\
$\nu_x$ (THz) & 1.30 & 1.25 & 1.36 & 1.13 \\
\vline width0pt height0pt depth5pt $\nu_y$ (THz) & 2.15 & 2.82 & 2.08 & 2.14 \\
\hline
\end{tabular}

}\footnotesize
\medskip
$\Delta E_{\rm d}$ is the
barrier for diffusion of the fullerenes along the nanotube axis,
$\Delta E_{\rm r}$ is the barrier for rotation of the fullerenes
about the nanotube axis, $\nu_{\rm d}$ is the frequency of small
relative vibrations of the fullerenes along the nanotube axis,
$\nu_z$ is the frequency of small relative rotational vibrations of
the fullerenes about the nanotube axis, $\nu_x$ and $\nu_y$ are the
frequencies of small relative rotational vibrations of the
fullerenes about the two mutually perpendicular lateral axes.
\end{table}

Let us discuss the reason of the significant changes of the barriers
$\Delta E_{\rm d}$ and $\Delta E_{\rm r}$ at the Peierls transition
by the example of the barrier $\Delta E_{\rm r}$ for rotation of the
Fe@C$_{20}$ endofullerene about the nanotube axis.
Figure~\ref{fig:05} presents the dependences $E_{{\rm W}i}(\varphi)$
of the interaction energies between the endofullerene and individual
atoms of the nanotube on the angle $\varphi$ of rotation of the fullerene. Figure~\ref{fig:05} shows that maxima of dependences
$E_{{\rm W}i}(\varphi)$ for individual atoms of the nanotube occur at
different angles $\varphi_{{\rm m}i}$ and so the dependence $E_{\rm
W}(\varphi)$ of total energy on the angle of rotation is essentially
smoothed. In other words, the barrier $\Delta E_{\rm r}$ in the
dependence $E_{\rm W}(\varphi)$ of the {\it total} interaction
energy between the endofullerene and the nanotube is less by an
order of magnitude than the barriers $\Delta E_{{\rm r}i}$ in the
dependences of the interaction energy between the endofullerene and
{\it only one of the nanotube atoms}. Thus, the barrier $\Delta
E_{\rm r}$ is very sensitive to the values of the barriers $\Delta
E_{{\rm r}i}$ and angles $\varphi_{{\rm m}i}$. Therefore, the barrier
$\Delta E_{\rm r}$ changes considerably at the found changes of the
barriers $\Delta E_{{\rm r}i}$ and angles $\varphi_{{\rm m}i}$ obtained
from the Peierls distortions of the nanotube structure and the
change of the nanotube symmetry at the transition. It should also be noted that small barriers to relative motion of nanoobjects resulting from the compensation of contributions of individual atoms to the barriers
is a phenomenon well studied by the examples of such systems as
double-shell carbon nanoparticles~\cite{Lozovik00, Lozovik02},
double-walled carbon nanotubes~\cite{Kolmogorov00, Belikov04,
Lozovik03, Lozovik03a, Damnjanovic03, Damnjanovic02, Vukovic03} and
a graphene flake in a graphite surface~\cite{Matsushita05}. The
considerable changes of barriers for relative rotation of shells at
small changes of shell structure were found also for double-shell
carbon nanoparticles~\cite{Lozovik00, Lozovik02}.

The frequencies of small vibrations of the fullerenes along the
nanotube axis ($\nu_{\rm d}$), rotational vibrations about the
nanotube axis ($\nu_z$) and rotational vibrations about two mutually
perpendicular lateral axes ($\nu_x$, $\nu_y$) are also calculated
and listed in Table \ref{table1}. The most remarkable change of
frequency as a result of the structural phase transition corresponds
to rotational vibrations of the Fe@C$_{20}$ fullerene about the
nanotube axis (this agrees with the changes of the barriers).

\section{Dynamical behaviour of molecules inside nanotube}

Let us consider the possible changes of the dynamical behaviour of
the C$_{20}$ and Fe@C$_{20}$ fullerenes inside the (8,8) nanotube
caused by the structural phase transition. The Peierls instability
transition temperature $T_{\rm P}$ was estimated for the (5,5)
nanotube to correspond to temperature range $T_{\rm P} \simeq
1$--$15$~K \cite{Mintmire92, Sedeki00, Huang96}. Both barriers
$\Delta E_{\rm d}$ and $\Delta E_{\rm r}$ and the thermal energy
$k_{\rm B}T_{\rm P}$ are of the same order of magnitude at the
structural Peierls phase transition (see Table \ref{table1}).
Therefore, dramatic changes of the diffusion and drift over these
barriers can take place at the Peierls transition for the considered
pairs of the encapsulated molecules and the nanotube.

Recently, the diffusion and drift in the periodic potential surface
of the interaction energy dependent on the displacement $z$ along
the nanotube axis and the angle $\varphi$ of the rotation about the
nanotube axis were considered for the thread-like relative motion of
walls of double-walled carbon nanotubes~\cite{Lozovik03,
Lozovik03a}. The expressions for the diffusion coefficient and
mobility of a movable wall were obtained~\cite{Lozovik03,
Lozovik03a}. In this study, we consider specific cases of diffusion
of the molecules along the nanotube axis and rotational diffusion
about this axis. In this case, the expressions for diffusion
coefficients, $D_{\rm d}$ and $D_{\rm r}$, mentioned above and
corresponding to the diffusion along the nanotube axis and
rotational diffusion about this axis, respectively, take the form:
\begin{align}\label{d}
   D_{\rm d} &= \frac{\Omega_{\rm d} \delta_{\rm d}^2}{2} \exp
\left(-\frac{\Delta E_{\rm d}}{k_{\rm B} T} \right), \notag\\
   D_{\rm r} &= \frac{\Omega_{\rm r} \delta_{\rm r}^2}{2} \exp \left(-\frac{\Delta E_{\rm
r}}{k_{\rm B} T} \right),
\end{align}
where $\Omega_{\rm d}$ and $\Omega_{\rm r}$ are the pre-exponential
multipliers in the Arrhenius formula for the frequency of jumps of
the molecule between two neighbouring global minima of the potential
surface $E_{\rm W}(\varphi,z)$, $\delta_{\rm d}$ is the distance
between neighbouring global minima for the motion of the molecule
along the nanotube axis, $\delta_{\rm r}$ is the angle between
neighbouring global minima corresponding to the molecule rotation
about the nanotube axis and $k_{\rm B}$ is the Boltzmann constant.
The mobility $B_{\rm d}$ for the motion along the axis can be easily
obtained from the diffusion coefficient $D_{\rm d}$ using the
Einstein ratio $D_{\rm d}/B_{\rm d} = k_{\rm B} T$.
Figure~\ref{fig:03} shows that $\delta_{\rm d} = 0.123$~nm and
$\delta_{\rm d} = 0.37$~nm for the (8,8) nanotube with the structure
of the metallic phase and the Kekule structure, respectively, and
$\delta_{\rm r} = 22.5^\circ$ for the both structures of this nanotube.

The value of the pre-exponential multiplier $\Omega$ in the
Arrhenius formula is usually considered to be related with the
frequency $\nu$ of corresponding vibrations. We suppose that the
ratio $\Omega/\nu$ remains the same for relative motion of different
carbon nanoobjects with graphene-like structure (nanotube walls and
fullerenes). For reorientation of the fullerenes of the
C$_{60}$@C$_{240}$ nanoparticle, the frequency multiplier $\Omega$
was estimated by molecular dynamics simulations having the value of
$650\pm 350$~GHz \cite{Lozovik00, Lozovik02}. We expand the
potential surface of the intershell interaction energy near the
minimum using the same empirical potential as in~\cite{Lozovik00,
Lozovik02}, and calculate the frequencies of small relative
librations of the shells. The calculated libration frequency has the
value $\nu \approx 50$~GHz, an order of magnitude less than that of
the frequency multiplier $\Omega$. In the estimations of this study,
we use the values $\Omega_{\rm d}\approx10\nu_{\rm d}$ and
$\Omega_{\rm r}\approx10\nu_z$ for the pre-exponential multipliers.

\begin{figure}[!t]
\hfil\includegraphics{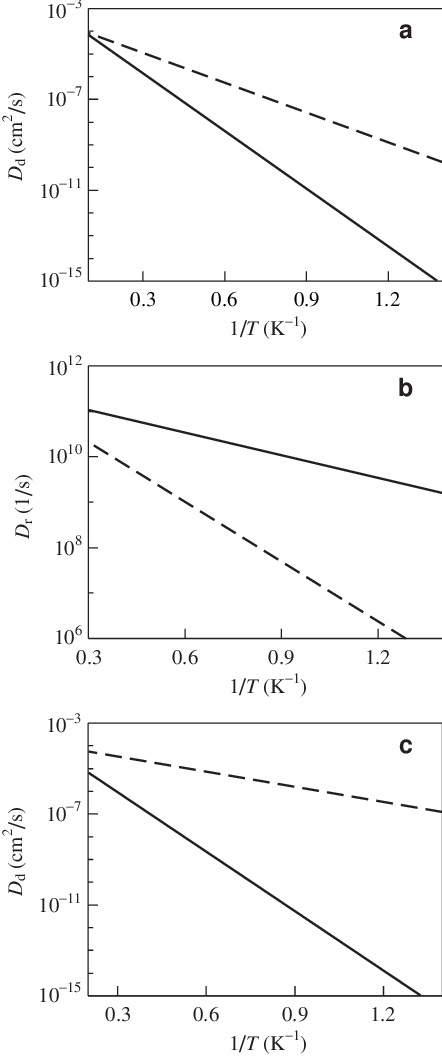}
\caption{{\bf Dependences of the diffusion coefficients on the reciprocal of temperature $1/T$.}
{\bf (a, c)} The diffusion
coefficient $D_{\rm d}$; {\bf (b)} the diffusion coefficient $D_{\rm
r}$. {\bf (a, b)} The C$_{20}$ fullerene; {\bf (c)} the Fe@C$_{20}$
endofullerene. Solid lines: The (8,8) carbon nanotube with the
Kekule structure; dashed lines: the (8,8) carbon nanotube with the
structure of metallic phase.}\label{fig:06}
\end{figure}

The temperature dependences of the diffusion coefficients, $D_{\rm
d}$ and $D_{\rm r}$, estimated using expressions (\ref{d}) are shown
in Figure~\ref{fig:06}. The dependence of the interaction energy
$E_{\rm W}(z)$ on the displacement of the fullerene C$_{20}$ has two
different barriers between the neighbour minima for the case of the
nanotube with the Kekule structure (see Figure~\ref{fig:04}a). Since
the frequency of the jumps of the fullerene between the neighbour
minima exponentially depends on the barrier, the contribution of
jumps over the lower barrier into the total diffusion coefficient is
disregarded in these estimations. The temperature range
corresponding to the Peierls transition temperature estimates
$T_{\rm P} \simeq 1$--$15$~K \cite{Mintmire92, Sedeki00, Huang96}
and the cases $\Delta E_{\rm d}/k_{\rm B} T < 1$ and $\Delta E_{\rm
r}/k_{\rm B} T < 1$, where the Arrhenius formula is adequate, are
considered. (For rotation of the Fe@C$_{20}$ inside the nanotube
with the Kekule structure the Arrhenius formula is not applicable
($\Delta E_{\rm r}/k_{\rm B} T
> 1$) at this temperature range; this case is considered below.) Figure~\ref{fig:06} shows that the changes of the diffusion coefficients, $D_{\rm d}$ and
$D_{\rm r}$, at the Peierls transition can be of orders of
magnitude. It is of interest that the diffusion coefficient $D_{\rm
r}$ for rotational diffusion of the C$_{20}$ fullerene decreases at
the Peierls transition with the increase of temperature.

If a molecule is encapsulated inside a nanotube without a structural
phase transition, the jump rotational diffusion takes place at low
temperatures, $\Delta E_{\rm r}/k_{\rm B} T > 1$, and the free
rotation of the molecule occurs at high temperature, $\Delta E_{\rm
r}/k_{\rm B} T < 1$. Orientational melting (a loss of the
orientational order with an increase of temperature) has a crossover
behaviour if the structural phase transition is absent. Firstly,
orientational melting was considered for two-dimensional clusters
with shell structure~\cite{Lozovik87, Lozovik90, Bedanov94,
Lozovik98} and later for double-shell carbon
nanoparticles~\cite{Lozovik00, Lozovik02}, double-walled carbon
nanotubes~\cite{Belikov04, Kwon98, Bichoutskaia06a} and carbon
nanotube bundles~\cite{Kwon00}. In the case where a molecule is
encapsulated inside a nanotube with a structural phase transition
and the barrier $\Delta E_{\rm r}$ for rotation of the molecule is
greater for the high-temperature phase than for the low-temperature
phase, an inverse orientational melting (a loss of the orientational
order with a decrease of temperature) is possible. In other words,
the inverse orientational melting takes place if the Peierls
transition temperature lies in the range $\Delta E_{\rm rl} < k_{\rm
B} T_{\rm P} < \Delta E_{\rm rh}$, where $\Delta E_{\rm rh}$ and
$\Delta E_{\rm rl}$ are the barriers for molecule rotation
corresponding to high-temperature and low-temperature phases of the
nanotube, respectively. For the considered molecules inside the
(8,8) nanotube, these temperature ranges are estimated to be $3.8 <
T_{\rm P}~\text{(K)} < 10$ and $0.58 < T_{\rm P}~\text{(K)} < 5.1$
for the C$_{20}$ and Fe@C$_{20}$ fullerenes, respectively (see Table
\ref{table1}). As these temperature ranges are in agreement with the
Peierls transition temperature estimates $T_{\rm P} \simeq
1$--$15$~K \cite{Mintmire92, Sedeki00, Huang96}, we predict that the
inverse orientational melting is possible for the systems
considered. The inverse orientational melting should be more
prominent for the case of the Fe@C$_{20}$ fullerene with the greater
ratio of the barriers $\Delta E_{\rm rh}/\Delta E_{\rm rl}$.

Let us discuss the possibility of observing the changes of the
dynamical behaviour of molecules inside armchair carbon nanotubes at
the Peierls transition. We believe that the most promising method is
high-resolution transmission electron microscopy. This method was
used for visualizing dynamics of processes inside nanotubes, such as
reactions of fullerene dimerization with monitoring of
time-dependent changes in the atomic positions~\cite{Koshino10} and
rotation of fullerene chains~\cite{Warner08}. The rotational
dynamics of C$_{60}$ fullerenes inside carbon nanotube was studied
also by analysing the intermediate frequency mode lattice vibrations
using near-infrared Raman spectroscopy~\cite{Zou09}. The
orientational melting in a single nanoparticle may be revealed also
by IR or Raman study of the temperature dependence of width of
spectral lines. A specific heat anomaly in multiwalled carbon
nanotubes may be caused by the orientational order--disorder
transition~\cite{Jorge10}. In the case of encapsulated magnetic
molecules (for example, the Fe@C$_{20}$ endofullerene considered
above), the study of the temperature dependence of the electron spin
resonance spectra could yield information on the molecule rotational
dynamics of these molecules~\cite{Simon06}.

%%%%%%%%%%%%%%%%%%%%%%
\section{Conclusive remarks}
In this Letter, we consider the changes of dynamical behaviour of
fullerenes encapsulated in armchair carbon nanotubes caused by the
Peierls transition in the nanotube structure by the example of the
C$_{20}$ and Fe@C$_{20}$ fullerenes inside the (8,8) nanotube. We
apply the DFT approach to calculate the structure of the C$_{20}$
and Fe@C$_{20}$ fullerenes. The ground state of the (8,8) nanotube
is found to be the Kekule structure using the method of molecular
orbitals. The Lennard-\hspace{0pt}Jones potential is used for
calculating the barriers for motion of the fullerenes along the
axis and rotation about the axis of the (8,8) nanotube with the
Kekule structure and the structure with all equal bonds
corresponding to low-temperature and high-temperature phases,
respectively. We show that the changes in the coefficients of
diffusion of the fullerenes along the nanotube axis and their
rotational diffusion at the Peierls transition can be as much as
several orders of magnitude. The possibility of the inverse
orientational melting at the Peierls transition is predicted. The
analogous changes of dynamical behaviour are also possible for other
large molecules inside armchair nanotubes. We believe that the
predicted dynamical phenomena can be observed using high-resolution
transmission electron microscopy, near-infrared Raman spectroscopy,
specific heat measurements, and by study of electron spin resonance
spectra for magnetic molecules.

%%%%%%%%%%%%%%%%%%
%\section*{Methods}
%  \subsection*{Methods sub-heading for this section}
%    Text for this sub-section \ldots
%
%  \subsection*{Another methods sub-heading for this section}
%    Text for this sub-section \ldots
%
%  \subsection*{Yet another sub-heading for this section}
%    Text for this sub-section \ldots

\section*{Abbreviation}

DFT, density functional theory.

%%%%%%%%%%%%%%%%%%%%%%%%%%%
\section*{Acknowledgements}
\small%  \ifthenelse{\boolean{publ}}{\small}{}
This work has been partially supported by the RFBR (Grants
08-02-00685 and 08-02-90049-Bel) and BFBR (Grant Nos.~F10R-062, F11V-001).
The atomistic calculations are performed on the SKIF MSU Chebyshev supercomputer and on the MVS-100K supercomputer at the Joint Supercomputer Center of the Russian Academy of Sciences.

%%  Bmc_article.bst  will be used to                       %%
%%  create a .BBL file for submission, which includes      %%
%%  XML structured for BMC.                                %%
%%  After submission of the .TEX file,                     %%
%%  you will be prompted to submit your .BBL file.         %%

{\footnotesize%\ifthenelse{\boolean{publ}}{\footnotesize}{\small}
 \bibliographystyle{bmc_article}  % Style BST file
  %\bibliography{poklonski}      % Bibliography file (usually '*.bib' ) 

  }

%%%%%%%%%%%

%\ifthenelse{\boolean{publ}}{\end{multicols}}{}

%%%%%%%%%%%%%%%%%%%%%%%%%%%%%%%%%%%
%%                               %%
%% Figures                       %%
%%                               %%
%% NB: this is for captions and  %%
%% Titles. All graphics must be  %%
%% submitted separately and NOT  %%
%% included in the Tex document  %%
%%                               %%
%%%%%%%%%%%%%%%%%%%%%%%%%%%%%%%%%%%

%%
%% Do not use \listoffigures as most will included as separate files

%%%%%%%%%%%%%%%%%%%%%%%%%%%%%%%%%%%
%%                               %%
%% Tables                        %%
%%                               %%
%%%%%%%%%%%%%%%%%%%%%%%%%%%%%%%%%%%

%% Use of \listoftables is discouraged.
%%

%%%%%%%%%%%%%%%%%%%%%%%%%%%%%%%%%%%
%%                               %%
%% Additional Files              %%
%%                               %%
%%%%%%%%%%%%%%%%%%%%%%%%%%%%%%%%%%%

%\section*{Additional Files}
%  \subsection*{Additional file 1 --- Sample additional file title}
%    Additional file descriptions text (including details of how to
%    view the file, if it is in a non-standard format or the file extension).  This might
%    refer to a multi-page table or a figure.

%  \subsection*{Additional file 2 --- Sample additional file title}
%    Additional file descriptions text.

%\end{bmcformat}
\end{document}